\documentclass[article,twocolumn,floats,nofootinbib,nobibnotes,superscriptaddress,10pt]{revtex4-1}

\usepackage{graphicx,amsmath,amssymb,hyperref}
\hypersetup{colorlinks=true, linkcolor=red, citecolor=magenta, urlcolor=blue}
\usepackage{dcolumn}
\usepackage{xcolor,colortbl}
\usepackage{enumitem}
\usepackage{eqnarray}
\usepackage{makecell}
\usepackage{cancel}
\usepackage{etoolbox}
\usepackage{comment}
\usepackage{amsfonts,mathtools}
\usepackage{tabularx,booktabs}
\usepackage{multirow}
\usepackage{rotating}

\begin{document}

\title{A New Boundary Condition on Reionization}

\newcommand\affSL{  
\affiliation{Department of Physics, Ben-Gurion University of the Negev, Be'er Sheva 84105, Israel}
}

\newcommand\affEK{
\affiliation{Department of Physics, Ben-Gurion University of the Negev, Be'er Sheva 84105, Israel}
}

\newcommand\affJM{
\affiliation{Department of Astronomy, The University of Texas at Austin,
2515 Speedway, Stop C1400, Austin, Texas 78712, USA}
\affiliation{Cosmic Frontier Center, The University of Texas at Austin, Austin, TX 78712}
}

\author{Sarah Libanore}
\email{libanore@bgu.ac.il}
\affSL

\author{Ely D. Kovetz}
\affEK

\author{Julian B. Mu\~noz}
\affJM

\author{Yonatan Sklansky}
\affJM

\author{Emilie Th\'elie}
\affJM

\begin{abstract}
The epoch of reionization (EoR) marks the last phase transition of hydrogen in our Universe, as it evolves from cold and neutral to hot and ionized in the intergalactic medium (IGM). While its endpoint and duration can be estimated from current observations, albeit with large uncertainties, there is no known avenue to constrain its onset. We propose a novel method based on the Pearson cross-correlation coefficient between 21-cm brightness temperature maps and line-intensity maps tracing star-formation (e.g., [OIII], CO, [CII]). This real-space estimator evolves from negative to positive as X-ray heating progresses, and saturates prior to the EoR. We predict a sharp drop from saturation during the earliest EoR stages, when the IGM ionized fraction is $x_{\rm HII}\lesssim 10\%$. 
We show that in standard scenarios, where IGM heating precedes reionization, the drop is a clear, model-robust signature that the EoR is still in its early stages, even when $\bar{x}_{\rm HII}$ cannot be measured precisely. This information is not accessible through the detection of an anticorrelation alone, which only indicates that reionization is ongoing. To assess the detectability of this feature, we provide a preliminary estimate of its signal-to-noise ratio in our fiducial scenario, assuming SPHEREx-{\it like} and SKAO-{\it like} noise levels, indicating that it is within reach of next-generation surveys.
The detection of the Pearson drop therefore will provide a unique anchor for the EoR onset, and an upper bound on $x_{\rm HII}$, complementing existing probes and tightening constraints on early galaxy formation models.
\end{abstract}

\maketitle

\section{Introduction}

How and when the Universe transitioned through the epoch of reionization (EoR,~\cite{choudhury2006physicscosmicreionization,2013_LoebFurlanetto,Wise_2019}) remains one of the most compelling open questions in modern cosmology. Driven by the very first galaxies, the EoR marks the time when their ultraviolet (UV) radiation~\cite{Miralda_1990,Madau_1991} ionized the neutral hydrogen in the intergalactic medium (IGM), representing the last major phase transition in cosmic history. Constraints on the EoR onset, duration, and scale-dependent evolution are crucial for understanding the properties of the first luminous structures, their underlying astrophysics, and our cosmological model.

So far, constraints on the EoR mainly tackle its end and approximate duration. The first evidence for a rising neutral hydrogen fraction, $\bar{x}_{\rm HI}$, came from studies of the Lyman-$\alpha$ (Ly$\alpha$) forest in quasar spectra. Before reionization ends, when $\bar{x}_{\rm HI}\gtrsim 10^{-3}$~\cite{choudhury2006physicscosmicreionization}, these absorption lines blend into the ``Gunn-Peterson trough"~\cite{Gunn_1965,Bahcall_1966}, observed up to redshift $z \sim 5.5$–6~\cite{Barkana:2001tx,Mesinger:2006kn,Mesinger_2010,McGreer:2014qwa,Davies:2017rxt,Greig:2018rts}, roughly a billion years after the Big Bang. As reviewed in,~e.g.,~Ref.~\cite{choudhury2006physicscosmicreionization} and references therein, from these spectra one can also constrain the sizes of ionized regions around quasars~\cite{Wyithe_2004,Wyithe_2005,Mesinger:2004gn}, 
the mean free path of ionizing photons~\cite{Zhu:2023ogg,Wolfson:2023puj}, 
and the slope of $\bar{x}_{\rm HI}(z)$ against $z$~\cite{Fan_2002,Fan_2006}. 
Together, these probes suggest that reionization is still ongoing at $z\in[5,6]$, with significant line-of-sight (LoS) variations.
Residual flux in quasar Ly$\alpha$ and Ly$\beta$ trough~\cite{Furlanetto_2005,Lidz:2005gn,McGreer:2011dm,Bosman:2021oom,Christenson:2021ldc,Qin:2021gkn} 
and damping wings in galaxy Ly$\alpha$ emission lines~\cite{Furlanetto:2004jz,Oh:2004rm,Mesinger:2007kd,becker2024dampingwingabsorptionassociated} 
further indicate LoS variations in the ionization state. 

Meanwhile, the duration of reionization can be estimated using cosmic microwave background (CMB) temperature and polarization data. Thomson scattering with free electrons damps small-scale temperature anisotropies, and produces large-scale polarization at $\ell \lesssim 30$~\cite{Hu:2001bc,challinor2004anisotropiescosmicmicrowavebackground,Challinor_2012}.
The amplitude of these fluctuations yields the present optical depth, $\tau$, an integrated measurement of the free-electron density. 
Within the standard cosmological model, $\Lambda$CDM, the optical depth $\tau$ has been constrained by WMAP~\cite{WMAP:2003elm} and  Planck~\cite{Planck:2018vyg} under the assumption of simple reionization models~\cite{adam_2016}.  
Allowing for more complex ionization histories or removing low-$\ell$ polarization data generally increases $\tau$~\cite{Heinrich_2017,
Delouis_2019,Qin:2020xrg,Heinrich_2021,Giare:2023ejv};  interestingly, recent works~\cite{craig2024nusgoodnews,Sailer:2025lxj,
Jhaveri:2025neg} suggest that a higher $\tau$ could alleviate the unphysical preference for negative $\sum m_\nu$ in DESI results~\cite{elbers2025constraintsneutrinophysicsdesi,elbers2025negativeneutrinomassesmirage,green2024cosmologicalpreferencenegativeneutrino}, as well as other cosmological anomalies, e.g.,~the Hubble tension~\cite{allali2025reionizationhubbleconstantcorrelations,allali2025cosmictauensionsindirectlycorrelate}. 

Further hints on the EoR timeline come from radio transients~\cite{Fialkov:2016fjb,Heimersheim:2021meu} and from the high-$z$ galaxies observed by JWST and HST. The decline of Ly$\alpha$ emitters signals a rising $\bar{x}_{\rm HI}$, as Ly$\alpha$ photons scatter in the neutral IGM with LoS variations~\cite{Furlanetto:2005ir,Mesinger:2007jr,Finkelstein:2019idd,Witstok_2025}. Meanwhile, from UV and H${\alpha}$ observations one can infer the ionizing photon budget and, in a model-dependent way, how fast the EoR proceeded~\cite{Barkana:2000ex,Barkana:2005pn,Robertson:2013bq,whitler2025zgtrsim9galaxy,chakraborty2025probingreionizationeragalaxiesjwst}. 
While these observables have traditionally pointed to the EoR starting at $z\sim 10$~\cite{Robertson:2015uda,Bouwens:2015vha,Finkelstein:2019sbd}, new JWST observations may prefer a higher ionizing budget, and thus an earlier start~\cite{Munoz:2024fas}.
Inferring the timing of reionization from these observables, however, requires modeling selection biases and extrapolating our models to fainter galaxies than currently observed. 
As such, at present they cannot precisely constrain the EoR onset. 

In this paper, we present a novel, model-independent method to constrain and uniquely identify the very beginning of the EoR, based on the cross-correlation of star-forming and reionization tracers measured with line intensity mapping (LIM,~\cite{Kovetz:2017agg,Bernal:2022jap}) surveys.

LIM is an emerging technique that measures all incoming photons in a given frequency range without resolving individual sources. Its observables are based on intensity fluctuations over large fields of view, which are treated statistically to probe the full galaxy and emitter population. 
A key branch of LIM targets the high-$z$ 21-cm line from the spin-flip transition of neutral hydrogen in the IGM~\cite{Pritchard:2011xb,Furlanetto:2006jb,Furlanetto:2015apc}. The 21-cm brightness temperature, 
\begin{equation}\label{eq:T21}
    T_{21}(\vec{x},z) = \frac{T_s(\vec{x},z)-T_{\rm CMB}(\vec{x},z)}{1+z}(1-e^{-\tau_{21}(\vec{x},z)}),
\end{equation}
is defined as the contrast between the CMB temperature $T_{\rm CMB}(z)$ and the spin temperature $T_s(z)$, 
weighted by the 21-cm optical depth $\tau_{21}(\vec{x},z)\propto x_{\rm HI}(\vec{x},z)$. 
The evolution of $T_{s}(\vec{x},z)$ depends on the relative abundance of the two hydrogen states, and follows the gas kinetic temperature after the first galaxies formed~\cite{Wouthuysen:1952AJ.....57R..31W,Field:1958PIRE...46..240F}, initially remaining lower than $T_{\rm CMB}$ due to the gas adibatically cooling. 
As X-rays from galaxies heat the gas, the 21-cm signal switches to emission, first in dense regions and then everywhere~\cite{Pritchard:2006sq,Mesinger:2012ys,Fialkov:2014kta,Munoz:2021psm}. X-rays, with their smaller cross section, travel far through the IGM before being absorbed, depositing energy over large scales; in contrast, UV photons that drive reionization are absorbed more locally, creating ionized bubbles where 21-cm emission vanished~\cite{Furlanetto:2004nh,Furlanetto:2004ha,Lin:2015bcw}. Both the 21-cm global signal $\bar{T}_{21}(z)$, and the 21-cm power spectrum $\Delta^2_{21}(k,z)$ are therefore sensitive to reionization, making the 21-cm line an ideal EoR probe. 
Moreover, cross correlating 21-cm and CMB refines the constraints on the EoR by detecting the patchy kinetic Sunyaev-Zel'dovich (kSZ) effect, which affects CMB temperature anisotropies through Doppler shifts due to electron scattering in ionized patches, and CMB polarization through the patchy screening related with ionization variations~\cite{Zahn:2005fn,McQuinn:2005ce,Dvorkin:2008tf,Ji:2021djj}. 

However, the detection of the 21-cm signal has proven more challenging than anticipated: global signal experiments such as EDGES~\cite{Bowman:2018yin} or SARAS~\cite{Singh:2021mxo}, suffer from strong foreground contamination and instrumental noise, while interferometers such as HERA~\cite{HERA:2021bsv,HERA:2021noe}, LOFAR~\cite{Mertens:2020llj,Ceccotti:2025bcd} and NenuFAR~\cite{Munshi:2023buw,Munshi:2025hgk} have so far only placed upper limits on the power spectrum. This situation is expected to improve dramatically with the commissioning of SKAO~\cite{SKA:2018ckk} in the next decade, which will enhance interferometric sensitivity and provide, for the first time, mapping of 21-cm fluctuations in real space.

Concurrently, LIM surveys have expanded to target emission lines from star-forming regions, whose intensities trace the underlying star formation rate (SFR). The observed specific intensity is given by
\begin{equation}\label{eq:Inu}
    I_\nu(\vec{x},z)=\frac{c}{4\pi\nu_{\rm rest}H(z)} \int dM_h\frac{dn}{dM_h}(\vec{x},z)L(\vec{x},z),
\end{equation}
where $\nu_{\rm rest}=(1+z)\nu$ is the rest frame frequency of the target, $c$ the speed of light, $H(z)$ the Hubble parameter. The intensity is obtained by integrating the local halo mass function (HMF, $dn/dM_h$) weighted by the line luminosity $L(\vec{x},z)$, typically a function of the SFR.  
Maps of UV and optical lines ([OII], [OIII], H$\alpha$, H$_\beta$) will soon be produced by SPHEREx~\cite{SPHEREx:2014bgr} up to $z\!\sim\! 9$ ($z\sim 12$ for [OII]). Meanwhile, COMAP~\cite{Keating:2020wlx} is observing the rotational CO line, and COMAP-wide~\cite{COMAP:2021nrp} will soon cover higher $z$ and wider areas in overlap with LOFAR, while TIME~\cite{TIME} and FYST~\cite{Karoumpis_2022} will follow the CONCERTO~\cite{CONCERTO2020} experiment in targeting the far-infrared [CII] line in the EoR.

Star-forming lines are not {\it per-se} direct probes of the EoR phase transition, but their information can be exploited through cross correlation with the 21-cm signal. Cross correlation mitigates noise, since foregrounds and systematics are uncorrelated across surveys~\cite{Furlanetto:2006pg,Lidz:2011dx,McBride_2024}. Moreover, as first proposed in Ref.~\cite{Lidz:2008ry}, the cross power spectrum between 21-cm and a SFR tracer such as LIM,
\vspace{-4mm}
\begin{equation}
    \Delta^2_{\rm cross}(k,z)=\frac{k^3}{2\pi^2}\langle\delta_{21}(k,z)\delta^*_\nu(k,z)\rangle,
\end{equation}
and the corresponding cross-correlation coefficient,
\vspace{-1mm}
\begin{equation}\label{eq:r_cross}
    r_{\rm cross}(k,z)=\frac{\Delta^2_{\rm cross}(k,z)}{\sqrt{\Delta^2_{21}(k,z)\Delta^2_\nu(k,z)}},
\end{equation}
are powerful indicators of the EoR process~\cite{Moriwaki:2019dbg,Hutter:2023rja,Moriwaki:2024kvp,Fronenberg_2024,Kannan:2021ucy,Sun:2024vhy,Chang:2019xgc,Beane_2019}. 

In fact, in regions with high SFR, line emission grows stronger, while the 21-cm signal decreases and eventually vanishes as $x_{\rm HI}$ (and thus $\tau_{21}$ in Eq.~\eqref{eq:T21}) approaches zero. Consequently, the SFR$\times T_{21}$ correlation is negative inside (partially) ionized regions. 
Building on this, Refs.~\cite{Moriwaki:2019dbg,Kannan:2021ucy} show that $r_{\rm cross}(k,z)$ between 21-cm and [OIII] at 5007\,\AA\, exhibits a zero crossing during the EoR; similar discussion can be found for 21-cm$\times$[CII] in Refs.~\cite{Gong_2012,Dumitru:2018tgh,Sun_2023}, and 21-cm$\times$Ly$\alpha$ in Refs.~\cite{Heneka_2017,Heneka:2021aey}, further highlighting how well LIM cross correlations can probe the evolution of the IGM in these stages. The transition from positive to negative correlation provides a clear indication that reionization is underway; however, it requires the EoR to be already advanced, with bubbles large enough to anti-correlate the signal. In order to tackle earliest stages, we need to focus on smaller scales at higher redshifts. 

In this work, we propose a new observable feature in the 21-cm-LIM cross-correlation that can constrain the very onset of the EoR, rather than its progression. We start from the real-space two-point correlation function,
\begin{equation}
    \xi_{\rm cross}(r,z)=\langle\delta_{21}(x,z)\delta_\nu(x+r,z)\rangle,
\end{equation} 
which preserves information from specific physical scales that is otherwise mixed in different Fourier modes of the power spectrum. 
This makes the signal easier to interpret and connects its redshift evolution to the bubble size distribution, as previously investigated for the cross correlation between the 21-cm signal and Ly$\alpha$~\cite{Hutter:2023rja}. 

\begin{figure*}[ht]
    \centering
    \includegraphics[width=\linewidth]{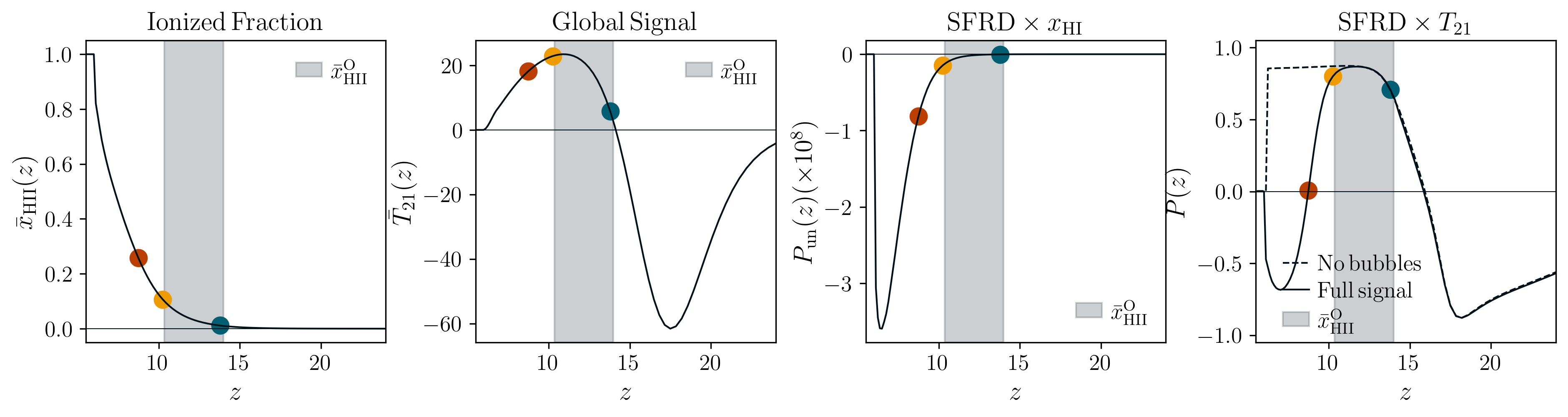}
    \vspace*{-.7cm}
    \caption{{\bf Left to right:} (i) average ionized fraction $\bar{x}_{\rm HII}=1-\bar{x}_{\rm HI}$; (ii) global signal $\bar{T}_{21}$; (iii) Pearson cross-correlation coefficient between the star formation rate density and the neutral fraction, ${\rm SFRD}\times x_{\rm HI}$ (unnormalized); (iv)  Pearson coefficient between the SFRD and the 21-cm brightness temperature, ${\rm SFRD}\times T_{21}$. 
    In all panels, parameters are fixed to their fiducial values in Fig.~\ref{fig:comparison} and the shaded region indicates the EoR onset, $\bar{x}^{\rm O}_{\rm HII} \leftrightarrow \bar{x}_{\rm HII}\in[1\%,10\%]$; Fig.~\ref{fig:scatter_plot} discusses how varying the astrophysical parameters affects this picture. The colored points keep track of the evolution.}
    \vspace*{-.3cm}
    \label{fig:SFRDxH}
\end{figure*}

In our case, we aim to probe the formation of the earliest, tiny ionized bubbles, which will manifest at the smallest scales possible. 
We therefore define the zero-lag (or one-point) real-space cross correlation estimator
\begin{equation}\label{eq:pearson}
\begin{aligned}
    P(z) &= \frac{\xi_{\rm cross}(r=0,z)}{\sqrt{\xi_{21}(r=0,z)\xi_\nu(r=0,z)}}\\
    &=\frac{\sum_i (T_{21,i}-\bar{T}_{21})(I_{\nu,i}-\bar{I}_\nu)}{\sqrt{\sum_i (T_{21,i}-\bar{T}_{21})^2\sum_i(I_{\nu,i}-\bar{I}_\nu)^2}},
\end{aligned}
\end{equation}
where the sums run over all voxels in the LIM maps.
Here, $T_{21,i}=T_{21}(\vec{x}_i,z)$ and $I_{\nu,i}=I_\nu(\vec{x}_i,z)$ denote the 21-cm brightness temperature and line intensity in voxel $i$, while $\bar{T}_{21}(z)$ and $\bar{I}_\nu(z)$ are their respective averages. Equation~\eqref{eq:pearson} measures the linear correlation between the two maps and is commonly known as the {\it Pearson coefficient}. 
In the following, we show that its redshift evolution encodes robust, model-independent features that track the onset of the EoR, defined here as the time when the ionized fraction $\bar{x}_{\rm HII}=1-\bar{x}_{\rm HI}$ is $\bar{x}_{\rm HII}\leq 10\%$. For our analysis, we use the publicly available codes \texttt{Zeus21}~\cite{Munoz:2023kkg,Cruz:2024fsv,Sklansky:2025} and~\texttt{oLIMpus}~\cite{Libanore:2025wtu}, to produce maps smoothed over $R=3\,{\rm Mpc}$ (unless otherwise stated), a scale well matched to the angular resolution and sensitivity of LIM experiments, e.g.,~COMAP~\cite{COMAP:2021nrp}. Ionized bubbles smaller than the resolution limit are modeled as partial reionization~\cite{Sklansky:2025}. 
While we defer a detailed study of the detectability of the EoR onset to future work~\cite{Libanore:2026}, towards the end of this paper we provide a preliminary estimate of the signal-to-noise ratio of the Pearson coefficient, to assess the feasibility of its detection given the noise levels of current and upcoming LIM surveys. We further expand our analysis in our follow-up paper~\cite{Thelie:2026nii}, where we investigate the two-point real-space cross correlation and its connection to the bubble mass function.

\section{Pearson Coefficient Physical Interpretation}

To build intuition for the physics underlying the Pearson coefficient, Fig.~\ref{fig:SFRDxH} shows the evolution of its key components. The first two panels display the average ionized fraction, $\bar{x}_{\rm HII}(z)$, and the global signal, $\bar{T}_{21}(z)$, while the third shows the unnormalized coefficient between star formation rate density (SFRD) and neutral fraction $x_{\rm HI}$.\footnote{In the third panel of Fig.~\ref{fig:SFRDxH}, we show the unnormalized Pearson coefficient $P_{\rm un}(z)=\sum_i(x_{{\rm HI},i}-\bar{x}_{\rm HI})({\rm SFRD}_i-\overline{\rm SFRD})$, to avoid numerical instabilities when $x_{\rm HI}$ approaches zero.}  

\noindent
The last panel shows the Pearson coefficient between SFRD and the 21-cm field, with (solid) or without (dashed) accounting for $x_{\rm HI}$ evolution. 

Once $T_{21}$ is in emission everywhere, the positive correlation between $T_{21}$ and SFRD saturates (blue dot). The saturation amplitude depends on the shot noise, which decorrelates the fields on small scales: for larger shot noise, the peak of $P(z)$ is farther below $1$. 

As the EoR begins, however, the Pearson coefficient drops quite rapidly: the departure from saturation cleanly marks the emergence of the first ionized bubbles, which are solely responsible for this deviation. The yellow dot highlights this point: in the SFRD$\times x_{\rm HI}$ panel, it anchors the moment when ionized regions become sufficient to drive a nonzero anticorrelation. In fact, before reionization, ${x}_{\rm HI}=1$ everywhere and $P_{\rm un}^{{\rm SFRD}\times x_{\rm HI}}\simeq 0$. 
Once the densest regions ionize, $P_{\rm un}^{{\rm SFRD}\times x_{\rm HI}}<0$. 
This transition propagates into SFRD$\times T_{21}$, forcing {\it $P(z)$ to fall from its saturation level}. Comparing with the first panel, we see that this occurs when $\bar{x}_{\rm HII}\leq 10\%$, which we label throughout as the {\it onset of reionization. 
This turnover in $P(z)$ provides a new way to detect it!}

While reaching the Pearson plateau is independent of the EoR onset (and its apparent alignment with $x_{\rm HII}\simeq 1\%$ in our fiducial model is coincidental),
the point where the Pearson estimator departs from saturation and begins to decline indeed traces the formation of the first ionized bubbles, probing the smallest scales that are not shot-noise dominated. 
As reionization proceeds, the amplitude of $P(z)$ continues to decrease, eventually turning negative (red dot). 
This corresponds to the situation highlighted in Refs.~\cite{Moriwaki:2019dbg,Kannan:2021ucy}, where the sign flip of the correlation was argued to indicate that reionization is underway. 
In contrast to the turnover from saturation we focus on in this paper, the zero crossing occurs later into reionization, when $\bar{x}_{\rm HII}\gtrsim 20\%$. 
Instead, our $P(z)$ estimator uniquely captures the earliest stages of the EoR.

Also, the SFRD$\times T_{21}$ estimator is more sensitive to the EoR evolution than the global signal itself, see second panel: the blue dot matches a still-rising $\bar{T}_{21}$; the $\bar{T}_{21}$ peak coincides here with $\bar{x}_{\rm HII}\!\sim \!10\%$; and the red dot with a still-large $\bar{T}_{21}$. Indeed, when the EoR begins, although $T_{21}(\vec{x},z)\!=\!0$ inside ionized regions, underdense IGM regions continue heating, raising their local $T_{21}$. 
These competing effects govern the $\bar{T}_{21}$ evolution and delay its null, making the exact position of these points parameter dependent. Unlike the $P(z)$ estimator, the global signal lacks sharp features solely related to the EoR onset.

\section{Dependence on the Astrophysical Parameters}

So far we have discussed the cross-correlation with the SFRD, which is cleaner from a modeling perspective. While the SFRD is not directly observable, similar conclusions hold for the cross-correlation of $T_{21}$ with an SFRD tracer. Since galaxy surveys are highly biased and impractical at such high redshifts, we propose instead to use LIM maps, where SFRD is traced by line-intensity fluctuations related with star-formation processes~\cite{Bernal:2022jap}.

\begin{figure}[th!]
    \hspace*{-.5cm}   
    \includegraphics[width=1.05\linewidth]{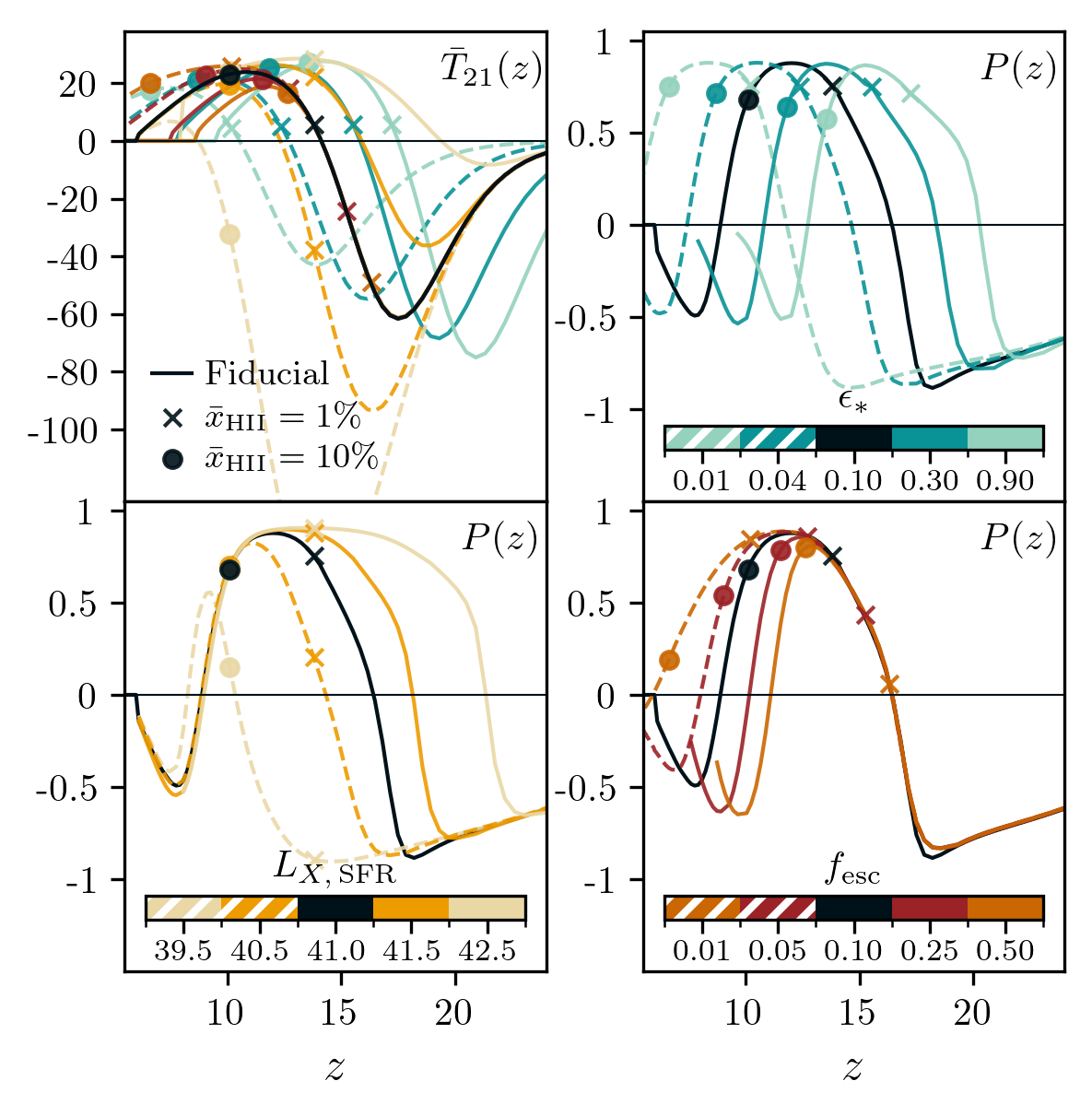}
    \vspace*{-.8cm}    
    \caption{Global 21-cm signal $\bar{T}_{21}(z)$ ({\bf top left panel}) and Pearson correlation coefficient $P(z)$ between the 21-cm and [OIII] maps, and their dependence on astrophysical parameters. In each curve, all parameters are fixed to their fiducial values except the one being varied. In the {\bf top right panel} (blue) we vary the star formation efficiency $\epsilon_*$; in the {\bf bottom left} (yellow) the X-ray luminosity $L_{X,\rm SFR}$; and in the {\bf bottom right} (red) the escape fraction $f_{\rm esc}$. Dashed lines indicate values smaller than the fiducial (which is always shown in black); lighter shades denote more extreme scenarios. Markers indicate the onset of the EoR, defined as $x_{\rm HII}\sim 1\%$-$10\%$ (``$\times$" and ``$\circ$" respectively). Same legend in the $\bar{T}_{21}$ panel. \vspace*{-.5cm}}
    \label{fig:comparison}
\end{figure}

In Fig.~\ref{fig:comparison}, we show an example in which the fiducial case (black line) is computed using a map of [OIII] at 5007\,\AA. We choose this line to build an analogy with Ref.~\cite{Moriwaki:2019dbg}; however, other lines could also be used, yielding similarly robust results at the theoretical level. A full analysis of detectability, however, depends on the specific instrument and LIM survey considered, as we discuss below.

To evaluate $L(\vec{x},z)$ in Eq.~\eqref{eq:Inu}, we adopt the model of Ref.~\cite{yang2024newframeworkismemission}.
The figure illustrates how $P(z)$ and the 21-cm global signal respond to variations in key astrophysical parameters. In particular, we consider $\epsilon_*$, the star formation efficiency; $L_{X,\rm SFR}$, the X-ray luminosity per unit SFR (in $\log_{10}$ units); and the escape fraction $f_{\rm esc}$ of ionizing photons into the IGM; for each parameter, we adopt a very broad range that encompasses not only the fiducial values but also extreme regimes. These parameters enter the $T_{21}$ and $I_\nu$ computations as detailed in Refs.~\cite{Munoz:2023kkg,Cruz:2024fsv,Libanore:2025wtu}.

Interestingly, across all scenarios but one, the drop from the Pearson plateau retains information on the EoR onset. Variations in $\epsilon_*$ shift the Pearson coefficient along the $z$-axis, but the drop in its amplitude consistently occurs at $\bar{x}_{\rm HII}\lesssim 10\%$. This reflects the fact that $\epsilon_*$ affects both the 21-cm and [OIII] emissions, ensuring that our estimator remains robust under such changes. 

\begin{figure}[th!]
    \hspace*{-.4cm}
    \includegraphics[width=1.05\linewidth]{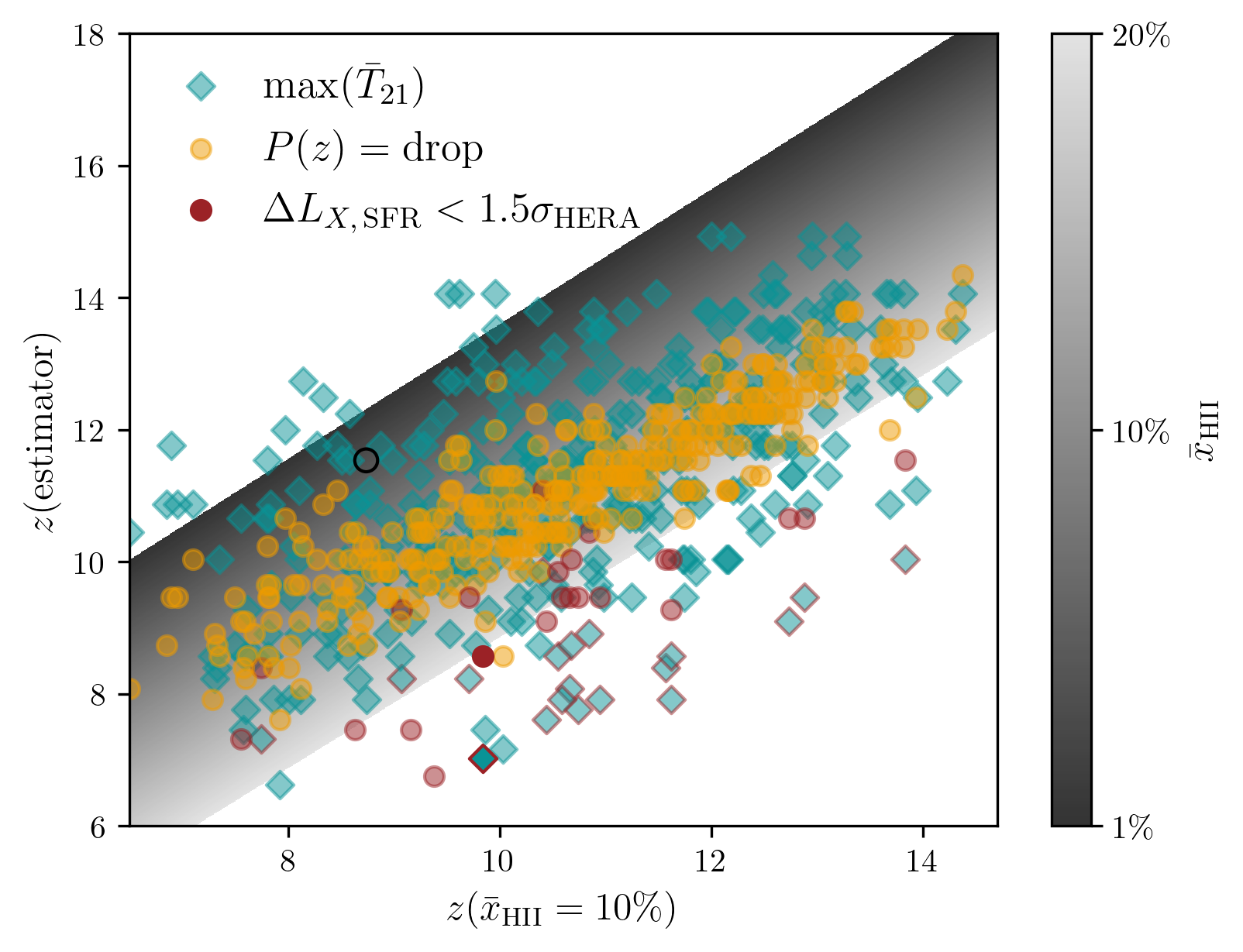}
    \vspace*{-.7cm}
    \caption{Each set of cyan/yellow points in the plot represents one of 350  realizations obtained using \texttt{oLIMpus} by randomly varying the astrophysical parameters, as discussed in the main text. For each model, the position in the plot is set by the redshift at which $\bar{x}_{\rm HII}=10\%$ ($x$-axis), and the redshift at which either the 21-cm global signal $\bar{T}_{21}(z)$ peaks (cyan squares), or the Pearson coefficient $P(z)$ drops by 10\% from its plateau (yellow dots). The red dots and red-contoured squares highlight models in which the sampled $L_{X,\,\rm SFR}$ is $1.5\sigma$ below the constraint in Ref.~\cite{Lazare:2023jkg}, i.e.,~the IGM is still cold when the EoR begins.
    The gray shaded region tracks the evolution of $\bar{x}_{\rm HII}$ from $1\%$ to $20\%$; $>\!70\%$ of the dots associated with the Pearson drop falls within $\bar{x}_{\rm HII}<10\%$ ; $>\!90\%$ at $\bar{x}_{\rm HII}<15\%$.
    The black, empty dot shows the position of the Pearson drop in the fiducial model previously discussed in the text.}
    \label{fig:scatter_plot}
\end{figure}

By contrast, $L_{X,\,\rm SFR}$ and $f_{\rm esc}$ influence only the evolution of $T_{21}$, and they are partially degenerate one with another as both can broaden (or narrow) the peak, extending (or shortening) the saturation plateau. This behavior reflects that $P(z)$ reaches saturation once the entire 21-cm map shows emission, $T_{21}>0$. 

This condition is met earlier when X-ray heating is highly efficient (yellow solid line). However, even for smaller values of $L_{X,\rm SFR}$, a comparable condition arises if the escape fraction is inefficient, as this delays the reionization progression (red dashed line). 
As long as the IGM is heated and the Pearson saturation is reached before reionization becomes significant, the drop in the $P(z)$ peak consistently occurs at $\bar{x}_{\rm HII}\lesssim 10\%$. The most extreme $f_{\rm esc}$ cases (light red lines) clearly show that our estimator is tied to an upper bound. When $f_{\rm esc}$ is very high (solid), the Pearson drop is sudden and almost immediate. Conversely, when $f_{\rm esc}$ is extremely small (dashed), the Pearson decrease is much slower; nevertheless, the departure from the peak occurs in both cases around $\bar{x}_{\mathrm{HII}} \sim 10\%$. Nevertheless, detecting this feature would still bound the redshift of the EoR onset. The only exception to this bound occurs for the lowest X-ray efficiency: $L_{X, \,\rm SFR}=39.5$, for which the drop occurs at $\bar{x}_{\rm HII}\gg10\%$. This corresponds to an extreme scenario in which reionization takes place in a still-cold IGM, as also the shape of the associated global signal shows. Although we chose to include this case for completeness, we note that low $L_{X, \,\rm SFR}$ values are disfavored by theoretical models~\cite{Fragos_2013}, as well as by observational limits on the 21-cm global signal~\cite{Singh:2017gtp,Bevins:2022clu} and power spectrum~\cite{HERA:2021noe,HERA:2022wmy} (though less robustly if mini halos that host PopIII stars are included~\cite{Lazare:2023jkg}). Therefore, this limiting case does not detract from the overall strength of our conclusions.

\section{Robustness Test}

We further test the robustness of our estimator through the scatter plot in Fig.~\ref{fig:scatter_plot}. We run \texttt{oLIMpus} multiple times, each time varying $\{\epsilon_*,f_{\rm esc},L_{X,{\rm SFR}}\}$. 
We define a multivariate Gaussian probability distribution, in which the center and standard deviation of each parameter are in agreement with HERA results~\cite{Lazare:2023jkg}. Note that this introduces a shift with respect to the fiducial model previously discussed in the text: while our fiducial choice assumes $\epsilon_*=0.1$, in agreement with the model in~Ref.~\cite{Munoz:2023kkg}, HERA results are centered around $\epsilon_*=0.03$ ($\log_{10}f_{*,10}=-1.25$ in Ref.~\cite{Lazare:2023jkg}).

For each run, we randomly sample the values of the parameters from the  distribution, and we produce the 21-cm and [OIII] maps. 
We measure the redshift where $\bar{x}_{\rm HII}< \{10\%,15\%,20\%\}$ and we compare it with the redshift of two different estimators: the peak of $\bar{T}_{21}$, and the drop of $P(z)$, defined as the point where the amplitude of this coefficient is $10\%$ smaller than the maximum level reached on the plateau. We verified that, among $350$ runs, in more than $70\%$ ($90\%$) of the cases the Pearson drop tracks $\bar{x}_{\rm HII}\leq 10\%$ ($15\%$). The remaining cases are mostly associated with scenarios in which $L_{X,\,\rm SFR}$ is much smaller than the fiducial value, so that the IGM is still cold at the onset of the EoR, before $P(z)$ reaches its plateau (similarly to the $L_{X,\rm SFR}=39.5$ case in Fig.~\ref{fig:comparison}). 

Figure~\ref{fig:scatter_plot} clearly shows that the Pearson drop has smaller scatter than the global signal maximum, and that its position tracks $\bar{x}_{\rm HII}\lesssim 10\%$ reliably. Therefore, our estimator provides complementary information and greater sensitivity to the EoR onset than the global signal. This is confirmed by Fig.~\ref{fig:comparison}, in which, in the $\bar{T}_{21}(z)$ panel, the markers for $\bar{x}_{\rm HII}=1\%$-$10\%$ fall on model-dependent sections of the curve: the $1\%$ marker lies on the rising edge of the global signal trough and can be either below or above zero depending on the model; the $10\%$ marker, although close to the $\bar{T}_{21}$ peak in most cases, shifts due to variations in $L_{X,\,\rm SFR}$ or $f_{\rm esc}$. Therefore, the EoR onset is difficult to pinpoint from the shape of $\bar{T}_{21}$, whereas the Pearson turnover provides a well-defined target. 

We verified that, for all models shown in Fig.~\ref{fig:scatter_plot}, the condition $P(z)=0$ is satisfied only at later stages of the EoR, corresponding to ionized fractions $\bar{x}_{\rm HII}>20\%$. Since the estimators previously discussed in the literature~\cite{Moriwaki:2019dbg,Kannan:2021ucy} rely on a change in the sign of the large-scale two-point correlation function—from positive to negative—this implies that they probe larger values of the mean ionized fraction $\bar{x}_{\rm HII}$ than the Pearson-correlation drop. Moreover, the detection of an anticorrelation is sensitive to an ongoing EoR but does not provide information on whether reionization has only recently begun. Combining these estimators with the Pearson coefficient therefore provides a more complete and powerful way to trace the evolution of the EoR.

On top of what is shown in Fig.~\ref{fig:scatter_plot}, we run further tests varying other astrophysical parameters, e.g.,~the $\alpha_*$ and $\beta_*$ parameters governing the halo-mass dependence of the SFR power law. Expanding the parameter set increases the scatter, but does not alter our conclusion: the large majority of the Pearson-drop points lie within $\bar{x}_{\rm HII}\leq 10\%$ and track $\bar{x}_{\rm HII}$ fractions smaller than the respective $P(z)=0$ and max$(\bar{T}_{21})$ estimators. Further investigation of this aspect will be presented in future work, where we will perform a comprehensive analysis of the impact of the various parameters involved~\cite{Libanore:2026}.

Furthermore, we verified this by both computing the two-point function and $P(z)$ on maps smoothed on larger scales than those shown in the plot (3\,Mpc): different smoothing scales probe bubbles of different characteristic sizes, suggesting that the real-space correlation function encodes information on the evolution of the bubble mass function. This goes beyond what is captured by the Pearson coefficient alone, and we will explore these additional aspects in upcoming work~\cite{Thelie:2026nii}. 

Finally, we tested that running a similar analysis on simulations by \texttt{21cmFASTv4}~\cite{Davies:2025wsa}, a new version of \texttt{21cmFAST}~\cite{Mesinger:2010ne}, leads to comparable conclusions: in both cases with and without mini halos that host PopIII stars, 
the Pearson coefficient shows saturation followed by a sharp drop at $\bar{x}_{\rm HII}\lesssim 10\%$. This confirms that our $\bar{x}_{\rm HII}$ bound conservatively defines the EoR onset and that the estimator is sensitive to very small ionized fractions. 

\section{Preliminary Signal-to-Noise Ratio}\label{sec:SNR}

As a final point of interest, we stress that applying our Pearson-drop estimator to real data will require a robust definition of the detectable departure from saturation, as well as incorporating instrumental noise and handling systematics in real space. This is more challenging than in Fourier space, where foregrounds can be mitigated by excising ‘wedge’ modes~\cite{Parsons:2012qh}, whereas in real space they directly affect the signal amplitude. 
Yet, it is not obvious whether Fourier-space estimators outperform real-space ones: computing the two-point function double-counts noise, while $P(z)$ requires only one-point functions,  
partially reducing noise propagation.
Additionally, current interferometers such as HERA and LOFAR cannot probe the smallest scales associated with the tiniest bubbles, whereas the Pearson coefficient can potentially still capture this information. 

To move beyond the noiseless case, it is necessary to define an estimator for the Pearson drop and to evaluate its SNR for 
realistic experimental configurations and state-of-the-art LIM surveys~\cite{Libanore:2026}. A comprehensive study is beyond the scope of this paper and left to future work.
Here, to provide a first indication of the feasibility of our observable beyond the noiseless regime, we present a preliminary estimate of the expected SNR of the Pearson coefficient in a simplified scenario, assuming noise levels comparable to those expected for current and upcoming LIM experiments. 
While this represents a useful assessment of the detectability of the Pearson observable itself, it does not yet establish the detectability of the drop from the saturation regime, which is left for future work.

We introduce the observed quantities 
\begin{equation}
    \begin{aligned}
        T_{21}^{\rm obs}(z)&=T_{21}(z)+n_{21},\\
        I_\nu^{\rm obs}(z)&=I_\nu(z)+n_\nu,
    \end{aligned}
\end{equation}
where $n_{21},\,n_\nu$ are the instrumental noises of the 21-cm and star-forming line maps. We assume that each of the noise terms fluctuates following a Gaussian distribution with zero mean and variance $\sigma_{{21}}^2,\,\tilde{\sigma}_{\nu}^2$ respectively, and that they are uncorrelated, $\langle n_{21}n_\nu\rangle=0$. With this in mind, we rewrite the Pearson coefficient, Eq.~\eqref{eq:pearson}, as 
\begin{equation}\label{eq:pearson_obs}
\begin{aligned}
    P^{\rm obs}(z)&=\frac{\sum_{i} \delta T_{21,i}^{\rm obs}\delta I_{\nu,i}^{\rm obs}}{\sqrt{\sum_{i}(\delta T_{21,i}^{\rm obs})^2\sum_i(\delta I_{\nu,i}^{\rm obs})^2}}\\
    &=\frac{{\rm Cov}(T_{21}^{\rm obs},I_\nu^{\rm obs})}{\sqrt{{\rm Var}(T_{21}^{\rm obs}){\rm Var}(I_\nu)^{\rm obs}}}\\
    &=\frac{{\rm Cov}(T_{21},I_\nu)}{\sqrt{[{\rm Var}(T_{21})+\sigma_{21}^2][{\rm Var}(I_\nu)^{\rm obs}+\tilde{\sigma}_\nu}^2]},\\
\end{aligned}
\end{equation}
where $\delta T_{21,i}^{\rm obs},\delta I_{\nu,i}^{\rm obs}$ denote the deviations of the observed quantities in the $i$-th voxel from their respective means, $\bar{T}_{21}^{\rm obs}=\bar{T}_{21},$ $\bar{I}_\nu^{\rm obs}=\bar{I}_\nu$; as in the original case, the Pearson coefficient is averaged over all voxels in the map. In the second line, we relied on the noise-decorrelation in the numerator, while for the denominator we separated the signal and noise contributions to the variance. Eq.~\eqref{eq:pearson_obs} represents the observed signal when the noise is non-negligible; the correlation here is attenuated relative to the noiseless case in Eq.~\eqref{eq:pearson}, and we can express
\begin{equation}
\begin{aligned}
    P^{\rm obs}(z) &= P(z)\sqrt{\frac{{\rm Var}(T_{21}){\rm Var}(I_\nu)}{[{\rm Var}(T_{21})+\sigma_{21}^2][{\rm Var}(I_\nu)^{\rm obs}+\tilde{\sigma}_\nu}^2]}\\
    &=P(z)\mathcal{D}(z),
\end{aligned}
\end{equation}
where $\mathcal{D}(z)$ is the damping factor. 

The noise associated with the Pearson measurement is 
\begin{equation}\label{eq:errP}
   \sigma_P(z) = \sqrt{\frac{(1-P^{\rm obs}(z)^2)^2}{N_{\rm eff}(z)}} ;
\end{equation}
the interested reader can find the detailed derivation in App.~\ref{app:noise_pearson}.
Finally, we can estimate 
\begin{equation}\label{eq:SNR}
    {\rm SNR}(z) = \frac{|P^{\rm obs}(z)|\sqrt{N_{\rm eff}(z)}}{1-P^{\rm obs}(z)^2} = \frac{|P(z)|\mathcal{D}(z)\sqrt{N_{\rm eff}(z)}}{1-[P(z)\mathcal{D}(z)]^2}.
\end{equation}
In the last line, we described the SNR in terms of: the noiseless $P(z)$ discussed in the rest of the paper; the damping factor $\mathcal{D}(z)$; and the number of independent modes measured by the surveys, $N_{\rm eff}(z)$. As described in App.~\ref{app:noise_pearson}, the latter can be estimated as 
\begin{equation}\label{eq:Neff}
    N_{\rm eff}(z)=\frac{\left[(1-P^{\rm obs}(z)^2)\sum_{k=k_{\rm min}^{\rm survey}}^{k_{\rm max}^{\rm survey}} N_{\rm modes}(k,z)\right]^2}{\sum_{k=k_{\rm min}^{\rm survey}}^{k_{\rm max}^{\rm survey}}N_{\rm modes}(k,z)(1-r_{\rm cross}^{\rm obs}(k,z)^2)^2},
\end{equation}
where $r_{\rm cross}^{\rm obs}(k,z)$ is the Fourier-space two-point correlation coefficient in Eq.~\eqref{eq:r_cross} computed on the $T_{21}^{\rm obs},I_\nu^{\rm obs}$ maps, while $N_{\rm modes}(k,z)$ depends on the volume observed by the two surveys (see Eq.~\eqref{eq:Nmodes}), 
\begin{equation}
    V_{\rm overlap}(z)=A_{\rm overlap}[{\rm rad}]\chi^2(z)\frac{c}{H(z)}\Delta z_{\rm overlap},
\end{equation}
$A_{\rm overlap}[{\rm rad}]$ and $\Delta z_{\rm overlap}$ being the overlapping area (expressed in radians) and the redshift bin covered by both the 21-cm and LIM surveys, $\chi(z)$ the comoving distance, $c$ the speed of light and $H(z)$ the Hubble factor. Finally, in Eq.~\eqref{eq:Neff} we set 
\begin{equation}
\begin{aligned}
    k_{\rm min}^{\rm survey}&={\rm max}(k_{\rm min}^\perp,k_{\rm min}^\parallel)\\
    &={\rm max}\left(\frac{\pi}{\chi(z)\sqrt{A_{\rm overlap}[{\rm rad}]}},\frac{\pi}{\Delta z_{\rm overlap}\,c/H(z)}\right)\\
    k_{\rm max}^{\rm survey}&={\rm min}(k_{\rm max}^\perp,k_{\rm max}^\parallel)\\
    &={\rm min}\left(\frac{\pi}{\chi(z)\theta[{\rm rad}]},\frac{\pi}{\delta z\,c/H(z)}\right), 
\end{aligned}
\end{equation}
where we chose $\theta [{\rm rad}]={\rm max}(\theta_{21},\theta_\nu)$, $\theta_{21,\nu}$ being the angular resolutions of the two surveys (expressed in radians), and $\delta z={\rm max}(\delta z_{21},\delta z_\nu)$ with  $\delta z_{21,\nu}=(1+z)/R_{21,\nu}$ the redshift resolutions, related with the spectral resolution of the surveys ($R_{21,\nu}=\lambda_{21,\nu}/\delta \lambda_{21,\nu}$). 

To provide an order-of-magnitude estimate of the number of voxels and the noise levels expected for current and upcoming LIM experiments, we assume an SKA-Low-{\it like} survey for the 21-cm maps and a SPHEREx-{\it like} survey for the star-forming line. For simplicity and tractability, we consider that the two surveys have an overlapping sky area of 
${A}_{\rm overlap}=200\,{\rm  deg}^2$, and an overlapping redshift range relevant for the onset and early evolution of the EoR, $z \in [6.5,12.5]$. This interval is divided into six bins of width $\Delta z_{\rm overlap}=1$.

Table~\ref{tab:survey} summarizes the adopted parameters; our choices are guided by results and discussions in the literature. For the 21-cm maps, the parameters in Tab.~\ref{tab:survey} are comparable to those assumed for the SKA-Low Deep Survey in Ref.~\cite{Koopmans:2015sua}. That work assumes that calibration and foreground subtraction are sufficiently accurate for their residuals to be subdominant with respect to thermal noise, and we adopt the same assumption here.
For the star-forming line, instead, we assume properties comparable to the specifications of SPHEREx Band 5 in Ref.~\cite{Bock:2025ijl}.\footnote{See also \url{https://github.com/SPHEREx/Public-products}.} Since the SPHEREx noise is quoted for angular pixels much smaller than the effective SKA-Low resolution, we rescale the noise level to the larger angular scale assuming uncorrelated noise, i.e.,${\tilde\sigma}_\nu=\sigma_\nu\,\theta_{\rm min}/\theta_{\rm max}$, where in our case $\theta_{\rm min}$ corresponds to the SPHEREx-{\it like} angular resolution, while $\theta_{\rm max}$ corresponds to the SKA-Low-{\it like} resolution. 

\begin{table}[ht!]
    \centering
    \begin{tabular}{|c|cccccc|}
    \hline
     & $z_{\rm min}$  & 
     $z_{\rm max}$  & 
     $\theta_i$ & 
     $R_i$    & 
     $\sigma_{i}$ & $\tilde{\sigma}_i^2=(\frac{\sigma_i{\theta_{\rm min}}}{\theta_{\rm max}})^2$ \\
    \hline
    SPHEREx-{\it like} & $6.5$  & 
     $12.5$  & 
     $5''$ & 
     $120$    & 
     $1800\,\frac{\rm Jy}{\rm sr}$ & $900\,\frac{\rm Jy^2}{\rm sr^2}$ \\
     \hline
     SKA-Low-{\it like} & $6.5$  & 
     $12.5$  & 
     $5'$ & 
     $1000$    & 
     $1$\,mK & $1\,$mK$^2$ \\
    \hline
    \end{tabular}
    \caption{Survey specifications for the two surveys assumed in our preliminary forecasts. From left to right, the table lists the minimum and maximum redshift observed, the angular and spectral resolution, the noise std, and the effective noise variance adopted in the analysis. Our choices are guided by comparisons with previous discussions in the literature and correspond to parameters broadly comparable to those of SKA-Low–{\it like} and SPHEREx–{\it like} surveys.
    }
    \label{tab:survey}
\end{table}

\begin{figure}
    \centering
    \includegraphics[width=\linewidth]{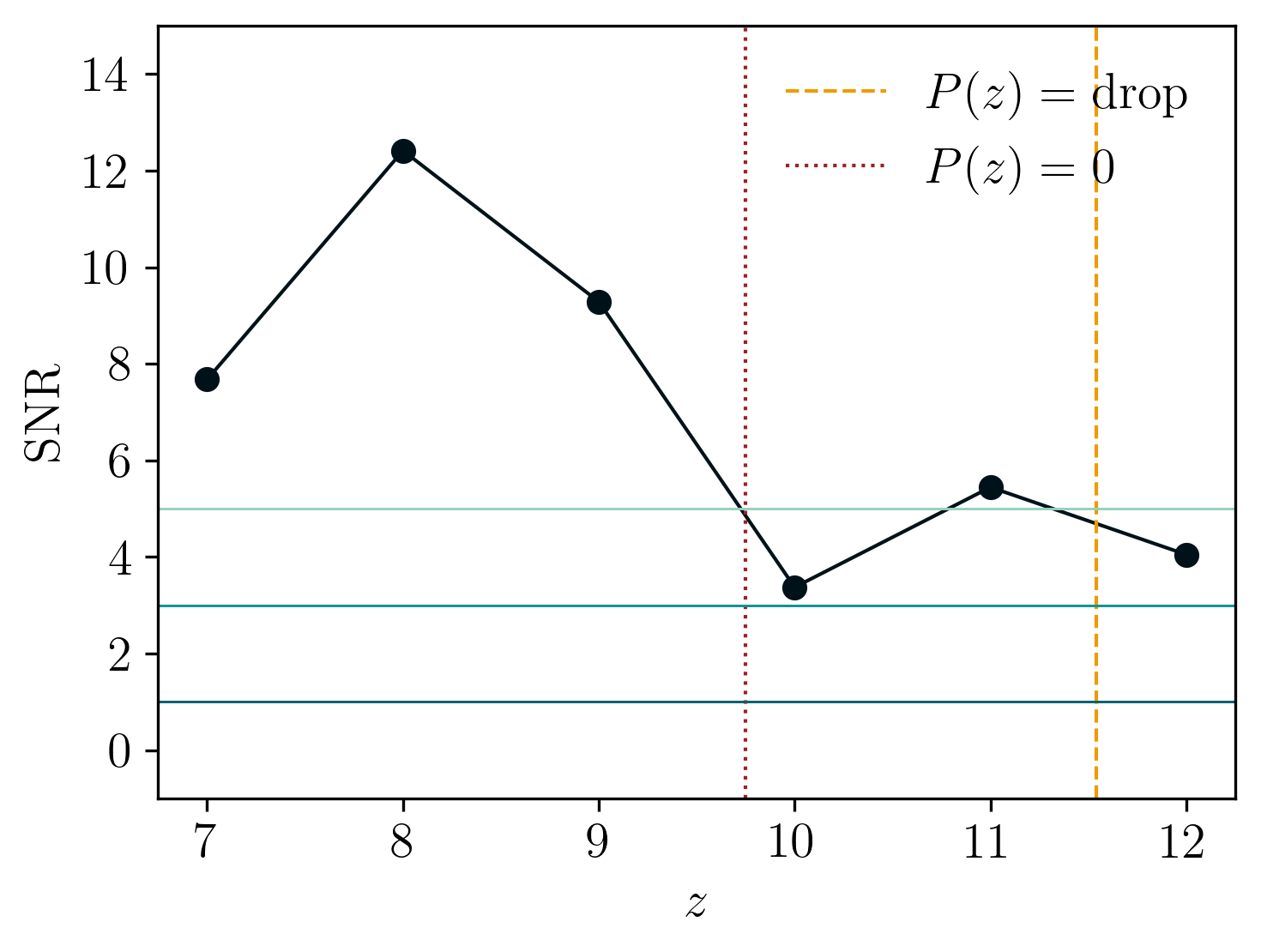}\vspace*{-.3cm}
    \caption{Preliminary estimate of the SNR of the Pearson coefficient for the cross correlation of 21-cm and star-forming line maps assuming a SKA-Low-{\it like} and a SPHEREx-{\it like} survey. The points show the position of the redshift bins, each having width $\Delta z =1$. The vertical lines indicate the position of the Pearson drop (dashed, marking the beginning of reionization) and of the zero-crossing (dotted). The horizontal lines indicate where the SNR reaches \{1,3,5\}.}
    \label{fig:SNR}
\end{figure}

Figure~\ref{fig:SNR} shows the SNR in Eq.~\eqref{eq:SNR} as a function of redshift for the fiducial model adopted in this paper. Over the full redshift range, the SNR reaches values high enough to allow a detection of the Pearson coefficient. 

Even if the SPHEREx-{\it like} survey did not extend to the high redshifts at which the Pearson drop occurs,~e.g.,~in scenarios where the EoR begins at very high $z$, the sufficiently large SNR at lower redshifts would still allow us to place meaningful constraints on the onset of the EoR. In particular, the SNR at low $z$ could be high enough to establish that the drop occurs at 
$z>z_{\rm max}$, implying $x_{\rm HII}(z=z_{\rm max})\geq 10\%$. Such a constraint would already be sufficient to rule out part of the currently allowed astrophysical parameter space.

\vspace*{-.3cm}
\section{Conclusions}
\vspace*{-.2cm}

In this paper we proposed a novel method to place a boundary condition on the start of the EoR using the Pearson correlation coefficient between 21-cm maps and LIM star-formation tracers (e.g., [OIII], [OII], CO, [CII]).

In all models consistent with current observations, this zero-lag real-space correlation exhibits a turnover associated with the formation of the very first bubbles. 

We demonstrated that the drop in the Pearson coefficient can be used to robustly bound the ionized fraction to be of order $\sim\!10\%$, while its redshift evolution can be used to track the evolution of the EoR. We provided a first estimate of the signal-to-noise ratio for the Pearson coefficient that could be achieved by cross-correlating LIM surveys with noise levels comparable to those expected for current and upcoming experiments, highlighting the potential of this estimator for astrophysical studies. 
Although this was obtained within a simplified framework neglecting 21-cm foreground contamination (making it potentially optimistic, even though in agreement with Ref.~\cite{Koopmans:2015sua}), the order of magnitude of the predicted SNR is promising. The detectability of the departure from the saturation regime and the subsequent estimate of $\bar{x}_{\rm HII}$ is left for future work, which will include a more detailed treatment of instrumental noise and foregrounds.
In summary, our estimator provides the first direct, model-independent method to constrain the onset of reionization, enabling coordinated LIM and 21-cm observations to probe its very beginning.


\vspace*{-.5cm}
\begin{acknowledgments}
\vspace*{-.2cm}

We thank the anonymous referees for their careful reading and insightful comments, which helped us improve the analysis and clarity of the manuscript.

We sincerely thank Adam Lidz and Jordan Mirocha for valuable discussions, and the anonymous referees for their useful comments.
SL acknowledges support by an Azrieli International Postdoctoral Fellowship. EDK acknowledges joint support from the U.S.-Israel Bi-national Science Foundation (BSF, grant No.\,2022743) and the U.S.\ National Science Foundation (NSF, grant No.\,2307354), as well as support from the ISF-NSFC joint research program (grant No.\,3156/23). 
JBM acknowledges support from NSF Grants AST-2307354 and AST-2408637, and the CosmicAI institute AST-2421782.

SL thanks UPenn, Philadelphia, and the University of Texas at Austin for hospitality during the last stages of this work; the visit was funded by the BGU/Philadelphia academic bridge grant, supported by the
Sutnick/Rosen Families Endowment Fund, Alton Sutnick/Helen Bershad BGU/Philadelphia Arts Collaboration Fund, The Alton Sutnick and Bettyruth Walter BGU/Philadelphia Collaboration Endowment Fund, The Mona Reidenberg Sutnick Memorial/Paula Bursztyn Goldberg Philadelphia Collaboration Endowment Fund, and the Sutnick/Zipkin BGU/Philadelphia Collaboration Expansion Fund. EDK thanks the Weinberg Institute at the University of Texas at Austin for warm hospitality during a sabbatical visit.
\end{acknowledgments}


\vspace*{-.5cm}
\appendix
\section{Pearson Coefficient Noise Estimate}\label{app:noise_pearson}
\vspace*{-.2cm}

In this Appendix, we derive the expression for the noise associated with the Pearson coefficient,~i.e.,~its variance. We begin by considering a pair of random variables $(X,Y)$ drawn $i=1,...N$ times from a bivariate normal distribution with  correlation 
\begin{equation}
    \rho = \frac{{\rm Cov}(X,Y)}{\sqrt{{\rm Var}(X){\rm Var(Y)}}}
\end{equation}
where ${\rm Var}(X)$ and ${\rm Var}(Y)$ are the variances of the two variables, and ${\rm Cov}(X,Y)$ is their covariance.
The sampled Pearson coefficient, in this case, reads as 
\begin{equation}\label{eq:pearson_stat}
    P = \frac{\sum_i (X_i-\bar{X})(Y_i-\bar{Y})}{\sqrt{\sum_i (X_i-\bar{X})^2\sum_i(Y_i-\bar{Y})^2}},
\end{equation}
and $P$ is an estimator of the population correlation $\rho$.
 
The distribution of the sampled Pearson coefficient is highly skewed; therefore, computing its variance requires introducing an approximation. One possible approach is the delta method (see, e.g., \cite{vanderVaart1998}), which approximates the distribution by expanding the estimator in terms of its derivatives with respect to ${\rm Var}(X)$, ${\rm Var}(Y)$, ${\rm Cov}(X,Y)$. However, because Eq.~\eqref{eq:pearson_stat} is highly non-linear, this approach requires at least a second-order expansion, which makes the computation cumbersome. 

Instead, we rely on the Fisher z-transform~\cite{Fisher1921}
\begin{equation}
    {\text z} = {\rm tanh}^{-1}(P).
\end{equation}
Crucially, when $N$ is large the z-distribution is well approximated by a normal distribution having mean $\langle {\rm z}\rangle={\rm tanh}^{-1}(\rho)$ and variance ${\rm Var}({\rm z})=1/(N-3)$.

\noindent
We can therefore expand the estimator around the mean,
using the partial derivative
\begin{equation}
    \frac{dP}{d{\rm z}}\biggl|_{\rm z=\langle z\rangle}=1-{\rm tanh}^{2}(\langle {\rm z}\rangle)=1-\rho^2. 
\end{equation}
Finally, using the standard error propagation and assuming $N \gg 3$, we obtain
\begin{equation}\label{eq:varP}
        {\rm Var}(P)=\left(\frac{dP}{d\rm z}\right)^2 {\rm Var}({\rm z})\simeq \frac{(1-\rho^2)^2}{N}.
\end{equation}

To apply this result to the Pearson coefficient in the main text, we now identify 
\begin{equation}
        X_{i}=T_{21,i}+n_{21,i},\quad
        Y_{i}=I_{\nu}+n_{\nu,i},
\end{equation}
where we neglected the redshift dependence for clarity.
Here, as in the main text, the noise terms are uncorrelated Gaussians with variances $\sigma_{21}^2, \sigma_\nu^2$; therefore, 
\begin{equation}
\begin{aligned}
    {\rm Cov}(X,Y)=&\,\,{\rm Cov}(T_{21}I_\nu),\\
    {\rm Var}(X)={\rm Var}(T_{21}) + \sigma_{21}^2,\quad& {\rm Var}(Y)= {\rm Var}(I_\nu) + \sigma_{\nu}^2,
\end{aligned}
\end{equation}
which leads to 
\begin{equation}
    \rho = \frac{{\rm Cov}(T_{21}I_\nu)}{\sqrt{({\rm Var}(T_{21}) + \sigma_{21}^2)({\rm Var}( I_\nu) + \sigma_{\nu}^2)}}=P^{\rm obs}.
\end{equation}

Identifying what $N$ represents, however, is more subtle. Equation~\eqref{eq:varP} is derived under the assumption that the 
$N$ measurements are statistically independent. This condition is not satisfied when estimating the Pearson coefficient from the voxels of a line-intensity map, since neighboring voxels are correlated both by the underlying cosmological signal and by instrumental effects such as beam smoothing and finite spectral resolution.

To obtain an order-of-magnitude estimate of the number of independent measurements entering the estimator, we instead consider the number of independent Fourier modes accessible to the survey. For a given wavenumber bin centered at 
$k$, this number is
\begin{equation}\label{eq:Nmodes}
    N_{\rm modes}(k,z) =
{V_{\rm overlap}(z)\,k^2 \Delta k}/({2\pi^2}),
\end{equation}
where $V_{\rm overlap}(z), k,\Delta k$ are determined by the survey properties.
These are the modes probed by the two-point estimator 
$r^{\rm obs}_{\rm cross}(k,z)$, which is measured from the same maps $X_i,Y_i$ used to compute the Pearson coefficient $P^{\rm obs}(z)$. Assuming that the Fourier modes are approximately independent and that the fields are close to Gaussian on the scales considered, the variance of the $r_{\rm cross}^{\rm obs}(k,z)$ estimator in a given $k$-bin can be written as
\begin{equation}\label{eq:Var}
{\rm Var}\left(r^{\rm obs}_{\rm cross}(k,z)\right)=
{\left(1-r^{\rm obs}_{\rm cross}(k,z)^2\right)^2}/
{N_{\rm modes}(k,z)}.
\end{equation}

This expression provides an estimate of the statistical uncertainty associated with each Fourier mode bin and can be used to infer an effective number of independent measurements contributing to the Pearson estimator. In fact, the Pearson coefficient can be approximately written as a mode-weighted average of $r_{\rm cross}^{\rm obs}(k,z)$ measured in the Fourier modes accessible to the survey,
\begin{equation}
P^{\rm obs}(z) \simeq
\frac{\sum_{k=k_{\rm min}^{\rm survey}}^{k_{\rm max}^{\rm survey}}
N_{\rm modes}(k,z) r^{\rm obs}_{\rm cross}(k,z)}
{\sum_{k=k_{\rm min}^{\rm survey}}^{k_{\rm max}^{\rm survey}}
N_{\rm modes}(k,z)} ,
\end{equation}
where the weights correspond to the number of independent modes contributing to each Fourier bin.
Under the assumption that the different Fourier bins are approximately independent, the variance of this estimator can be obtained by propagating the uncertainties of the individual mode estimates. Using Eq.~\eqref{eq:Nmodes}, this yields
\begin{equation}
\begin{aligned}
     \sigma_{P,\rm WA}(z)^2&=\frac{\sum_{k=k_{\rm min}^{\rm survey}}^{k_{\rm max}^{\rm survey}} N_{\rm modes}(k,z)^2{\rm Var}(r^{\rm obs}_{\rm cross}(k,z))}{[\sum_{k=k_{\rm min}^{\rm survey}}^{k_{\rm max}^{\rm survey}} N_{\rm modes}(k,z)]^2}\\
      &=\frac{\sum_{k=k_{\rm min}^{\rm survey}}^{k_{\rm max}^{\rm survey}} N_{\rm modes}(k,z)(1-r^{\rm obs}_{\rm cross}(k,z)^2)^2}{[\sum_{k=k_{\rm min}^{\rm survey}}^{k_{\rm max}^{\rm survey}} N_{\rm modes}(k)]^2}.
\end{aligned}
\end{equation}
By equating this expression with Eq.~\eqref{eq:varP}, we can write
\begin{equation}
    N_{\rm eff}(z) =\frac{\left[\left(1-P^{\rm obs}(z)^2\right)\sum_{k=k_{\rm min}^{\rm survey}}^{k_{\rm max}^{\rm survey}}N_{\rm modes}(k,z)\right]^2}{\sum_{k=k_{\rm min}^{\rm survey}}^{k_{\rm max}^{\rm survey}} N_{\rm modes}(k,z)(1-r^{\rm obs}_{\rm cross}(k,z)^2)^2},
\end{equation}
which is the expression adopted in the main text. 

To validate our analytical estimate, we also computed the variance of the Pearson estimator using a bootstrap approach. To this end, we first transformed the signal boxes $T_{21},I_\nu$ into noisy maps, $T_{21}^{\rm obs}, I_\nu^{\rm obs}$, by including the instrumental noise contribution. 

\bigskip
The procedure is the following:
\begin{enumerate}
\item We generate two noise boxes with the same size $L_{\rm box}$ and number of cells $N_{\rm box}$ as the signal maps. In each voxel we draw the noise value from the Gaussian distributions $n_{21}, n_\nu$.
\item The resulting noise maps are then added to the signal maps to obtain the noisy fields.
\item The noisy maps are smoothed with a Gaussian filter in order to mimic the angular resolution of the instruments. For each map, the Gaussian kernel is defined using a beam width $\sigma_{\rm beam} = \theta\chi(z)N_{\rm box}/[L_{\rm box}2\sqrt{2\log(2)}]$, where $\theta$ is the angular resolution of the survey with the largest beam, $\chi(z)$ the comoving distance at redshift, and $L_{\rm box}/N_{\rm box}$ the comoving size of a voxel.
\end{enumerate}
From these noisy maps\footnote{These maps are also used to measure $r_{\rm cross}^{\rm obs}(k,z)$ in Eq.~\eqref{eq:Neff}. } we construct the dataset of voxel pairs $
(T_{21,i}^{\rm obs},I_{\nu,i}^{\rm obs})$, which serves as the input for the bootstrap procedure. We generate $N_{\rm boot}=10^4$ bootstrap realizations of this dataset and estimate the variance of the Pearson coefficient from their dispersion.

\begin{figure}[hb!]
    \centering
    \includegraphics[width=\linewidth]{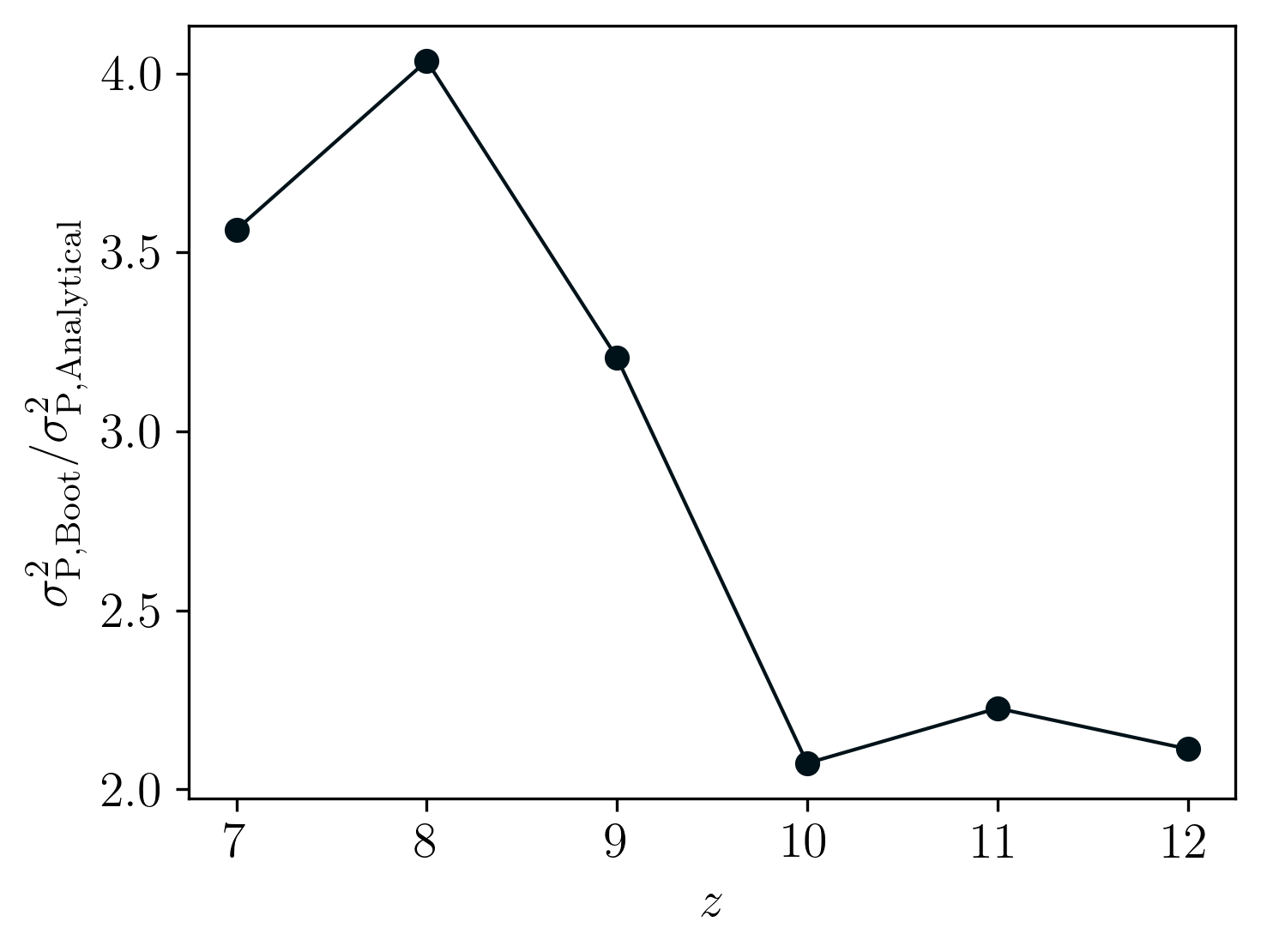}
    \caption{Ratio between the Pearson-estimator variance estimated from the bootstrap method $\sigma^2_{P,\rm Boot}$, and its analytical variance, $\sigma_{P,\rm Analytical}^2$. }
    \label{fig:compare_bootstrap}
\end{figure}

Figure~\ref{fig:compare_bootstrap} shows the ratio between the variance of the Pearson estimator computed using the bootstrap method and the analytical prediction from Eq.~\eqref{eq:errP}. The plot indicates that the two estimates are of the same order of magnitude, with the analytical variance only slightly underestimating the bootstrap result. Although both the analytical estimate and the bootstrap method rely on simplifying assumptions and have their own limitations, the level of agreement suggests that they provide a consistent and reasonable estimate of the uncertainty. A detailed treatment will be presented in future work~\cite{Libanore:2026}.

\newpage
\bibliography{biblio.bib}

\end{document}